\documentclass{article}

\def\giorno{30/7/2005}

\def\a{\alpha}
\def\b{\beta}

\def\de{\delta}   

\def\phi{\varphi}
\def\la{\lambda}
\def\s{\sigma}
\def\om{\omega}

\def\J{{\bf J}}
\def\P{{\cal P}}
\def\R{{\bf R}}
\def\T{{\rm T}}

\def\La{\Lambda}
\def\S{\Sigma}
\def\Om{\Omega}

\def\pa{\partial}
\def\xd{{\dot x}}

\def\grad{\nabla}     
\def\ss{\subset}
\def\sse{\subseteq}

\def\({\left(}
\def\){\right)}
\def\[{\left[}
\def\]{\right]}
\def\=#1{\bar #1}
\def\~#1{\widetilde #1}
\def\.#1{\dot #1}
\def\^#1{\widehat #1}
\def\"#1{\ddot #1}

\begin{document}

\title{\bf Finite group symmetry breaking}

\author{Giuseppe Gaeta \\
{\it Dipartimento di Matematica, Universit\`a di Milano} \\
{\it v. Saldini 50, I--20133 Milano (Italy)} \\ gaeta@mat.unimi.it
}

\date{\giorno}

\maketitle

\section{Introduction}

It is a commonplace situation that symmetric laws of
Nature give raise to physical states which are not symmetric. States related by symmetry operations are equivalent, but still Nature will select one of them.

As an example, consider a ferromagnetic system of interacting spins with no external magnetic field: the ``up'' and ``down'' states are equivalent, but one of the two will be chosen:
the interaction makes states with agreeing spin orientation (and
therefore macroscopic magnetization) energetically
preferred, and fluctuations will decide which state is actually
chosen by a given sample.

Finite group symmetry is also commonplace in Physics, in
particular through crystallographic groups occurring in condensed
matter physics -- but also through the inversions (C,P,T and their
combinations) occurring in high energy physics and field theory.

The breaking of finite groups symmetry has thus been thoroughly
studied, and general approaches exist to investigate it in
mathematically precise terms with physical counterparts. In
particular, a widely applicable approach is provided by the Landau
theory of phase transitions -- whose mathematical counterpart
resides in the realm of equivariant singularity and bifurcation
theory. In Landau theory, the state of a system is described by a finite dimensional variable (the {\it order parameter}), and physical states correspond to minima of a potential, invariant under a group.

In this article we describe the basics of symmetry breaking analysis for systems described by a symmetric polynomial; in particular we discuss generic symmetry breakings, i.e. those
determined by the symmetry properties themselves and independent
on the details of the polynomial describing a concrete system. We
also discuss how the plethora of invariant polynomials can be to
some extent reduced by means of changes of coordinates, i.e. how
one can reduce to consider certain types of polynomials with no
loss of generality. Finally, we will give some indications on
extension of this theory, i.e. on how one deals with symmetry
breakings for more general groups and/or more general physical
systems.

\section{Basic notions}

\subsubsection*{Finite groups}

A finite group $(G,\circ )$ is a finite set $G$ of elements $\{
g_0 , ... , g_N \}$ equipped with a composition law $\circ$, and
such that the following conditions hold: \par\noindent $(i)$ for
all $g,h \in G$ the composition $g \circ h$ belongs to $G$, i.e.
$g \circ h \in G$; \par\noindent $(ii)$ the composition is
associative, i.e. $(g \circ h) \circ k = g \circ (h \circ k )$ for
all $g,h,k \in G$; \par\noindent $(iii)$ there is an element in
$G$ -- which we will denote as $e$ -- which is the identity for
the action of $\circ$ on $G$, i.e. $e \circ g = g = g \circ e$ for
all $g \in G$; \par\noindent $(iv)$ for any $g \in G$ there is an
element $g^{-1}$ which is the inverse of $g$, i.e. $g^{-1} \circ g
= e = g \circ g^{-1}$.

In the following we omit the symbol $\circ$, i.e. we write $gh$ to
mean $g \circ h$. Similarly, we usually write simply $G$ for the
group, rather than $(G,\circ)$.

Given a subset $H \sse G$, this is a subgroup of $(G,\circ)$ if
$(H,\circ)$ satisfies the group axioms $(i)$--$(iv)$ above. Note
this implies that $e \in H$ whenever $H$ is a subgroup, and $\{ e
\}$ is a subgroup. Subgroups not coinciding with the whole $G$ and
with $\{ e \}$ are said to be {\it proper}.

Given two elements $g,h$ we say that $g h g^{-1} $ is the
conjugate of $h$ by $g$. The conjugate of a subgroup $H \sse G$ by
$g \in G$ is the subgroup of elements conjugated to elements of
$H$, $g H g^{-1} = \{ (g h g^{-1}) \ , \ h \in H \}$.

\subsubsection*{Group action}

In Physics, one is usually interested in a realization of an
abstract group as a group of transformations in some set $X$; in
physical applications, this is usually a (possibly, function) 
space or a manifold, and we refer to elements of $X$ as
``points''. That is, there is a map $\rho : G \mapsto End (X)$
from $G$ to the group of endomorphisms of $X$, such to preserve
the composition law:
$$ \rho (g) \cdot \rho (h) \ = \ \rho (g \circ h) \ \ \ \forall g,h \in G \ . $$
In this case we say that we have a {\it representation} of the abstract group $G$ acting in the {\it carrier} space or manifold $X$; we also say that $X$ is a $G$-space or $G$-manifold.
We often denote by the same letter the abstract element and its representation, i.e. write simply $g$
for $\rho (g)$ and $G$ for $\rho (G)$. \footnote{In many physically relevant cases, but not necessarily, $X$ has a linear structure and we consider linear endomorphisms. In this case we sometimes write $T_g$ for the linear operator representing $g$.}

If $x \in X$ is a point in $X$, the $G$-orbit $G(x)$ is the set of
points to which $x$ is mapped under $G$, i.e.
$$ G(x) \ = \ \{ y \in X \ : \ y = g x \ , \ g \in G \} \ \sse \ X \ . $$
Belonging to the same orbit is obviously an equivalence relation,
and partitions $X$ into equivalence classes. The {\it orbit space}
for the $G$ action on $X$, also denoted as $\Om = X/G$, is the set
of these equivalence classes. It corresponds, in physical terms,
to considering $X$ modulo identification of elements related by
the group action.

For any point $x \in X$, the {\it isotropy (sub)group} $G_x$ is
the set of elements leaving $x$ fixed,
$$ G_x \ = \ \{ g \in G \ : \ g x = x \} \ \sse \ G \ . $$
Points on the same $G$-orbit have conjugated isotropy subgroups:
indeed, $y = g x $ implies immediately that $G_y = g G_x g^{-1}$.

When a topology is defined on $X$, the problem arises if the 
$G$-action preserves it; if this is the case, we say that the
$G$-action is {\it regular}. In the case of a compact Lie group
(and {\it a fortiori} for a finite group) we are guaranteed the
action is regular.\footnote{A physically relevant example of
non-regular action is provided by the irrational flow on a torus.
In this case $G = R$, realized as the time $t$ irrational flow on the torus $X = {\bf T}^k$.}

\subsubsection*{Spontaneous symmetry breaking}

Let us now consider the case of physical systems whose state is
described by a point $x$ in the $G$-space or $G$-manifold $X$,
with $G$ a group acting by smooth mappings $g : X \to X$. In physical
problems, $G$ is quite often acting by linear and orthogonal
transformations\footnote{If this is not the case, the
Palais-Mostow theorem guarantees that, for suitable groups (including in particular finite ones) we can reduce to this case
upon embedding $X$ into a suitably larger carrier space $Y$.}.

Usually $G$ represents physical equivalence of states, and
$G$-orbits are collections of physically equivalent states. A
point which is $G$-invariant, i.e. such that $G_x = G$, is called
``symmetric'' for short.

Let $\Phi$ be a scalar function (potential) defined on $X$, $\Phi
: X \to \R$, possibly depending on some parameter $\mu$, such that
the physical state corresponds to critical points -- usually the
(local) minima -- of $\Phi$.

A concrete example is provided by the case where $\Phi$ is the
Gibbs free energy; more generally, this is the framework met in
the Landau theory of phase transitions (Landau 1937, Landau and
Lifshits 1958).

We are interested in the case where $\Phi$ is invariant under the
group action, or briefly $G$-invariant. That is, where
$$ \Phi (g x ) \ = \ \Phi (x) \ \ \ \forall x \in X \ , \ \forall
g \in G \ . \eqno(1) $$

A critical point $x$ such that $G_x = G$ is a {\it symmetrical
critical point}. If $G_x$ is strictly smaller than $G$, then $x$
is a {\it symmetry breaking critical point}.

If a physical system corresponds to a non symmetric critical
point, we have a {\it spontaneous symmetry breaking}: albeit the
physical laws (the potential function $\Phi$) are symmetric, the
physical state (the critical point for $\Phi$) breaks the symmetry
and chooses one of the $G$-equivalent critical points.

It follows from (1) that the gradient of $\Phi$ is covariant under
$G$. If $y = g (x)$, then the differential $(Dg)$ of the map $g :
X \to X$ is a linear map between the corresponding tangent spaces,
$(Dg) : \T_x X \to \T_y X$. The covariance amounts, with $\eta$
the Riemannian metric in $X$, to $ (\eta^{ij} \pa_j \Phi ) (g x) =
[ (Dg)^i_k \eta^{km} \pa_m \Phi ] (x)$; this is also written
compactly, with obvious notation, as\footnote{In the case of
euclidean spaces ($\eta = \delta$) and linear actions described by
matrices $T_g$, the covariance condition reduces to $(\grad
\Phi)^i (T_g x) = (T_g)^i_j [(\grad \Phi)^j (x)]$.}
$$ (\grad \Phi) (g x) \ = \ (Dg) \, [(\grad \Phi)(x)] \ . \eqno(2)  $$
As $(Dg)$ is a linear map, $(\grad \Phi)(x) = 0$ implies the
vanishing of $\grad \Phi$ at all points on the $G$-orbit of $x$.

We conclude that {\it critical points of a $G$-invariant potential
come in $G$-orbits}: if $x$ is a critical point
for $\Phi$, then all the $y \in G(x)$ are also critical points for
$\Phi$. We speak therefore of {\it critical orbits} for $\Phi$.

It is thus possible (thanks to the regularity of the $G$-action), and actually convenient, to study spontaneous symmetry breaking in the orbit space $\Omega = X/G$ rather than in the carrier manifold $X$ (Michel 1971).

If $G$ describes physical equivalence, physical states whose
symmetries are $G$-conjugated should be seen as physically
equivalent. An equivalence class of isotropy types under
conjugation will be said to be a {\it symmetry type}. We are thus
interested, given a $G$-invariant polynomial $\Phi$, to know the
symmetry types of its critical points. We denote symmetry types as
$[H] = \{ g H g^{-1} \}$, and say that $[H] < [K]$ if a group
conjugated to $H$ is strictly contained in a group conjugated to
$K$.

As we have seen, points on the same $G$-orbit have the same symmetry
type. On the other hand, points on different $G$-orbits can have the
same isotropy type (e.g., for the standard action of $O(n)$ in
$\R^n$, all collinear nonzero points will have the same isotropy
subgroup but will lie on distinct group orbits).

\section{G-invariant polynomials}

Consider a finite group $G$ acting in $X$.\footnote{Many of the notions and results mentioned in this section have a much wider range of applicability.} We look at the ring of $G$-invariant scalar
polynomials in $x^1,...,x^n$.

By the {\it Hilbert basis theorem}, there is a set $\{J_1 (x) ,
... , J_k (x) \}$ of $G$-invariant homogeneous polynomials of
degrees $\{d_1 , ... , d_k \}$ such that any $G$-invariant
polynomial $\Phi (x)$ can be written as a polynomial in the $\{
J_1 , ... , J_k \}$, i.e.
$$ \Phi (x) \ = \ \Psi \[ J_1 (x) , ... , J_k (x) \] \eqno(3) $$
with $\Psi$ a polynomial. (A similar theorem holds for smooth functions.)

The algebra of $G$-invariant polynomials is finitely generated, i.e. we can choose $k$ finite.
When the $J_a$ are chosen so that none of them can be written as a
polynomial of the others\footnote{Note that some of the $J_a$
could be written as non-polynomial functions of the others, and
the $J_\a$ could verify polynomial relations. E.g., consider the
group $Z_2$ acting in $\R^2$ via $g : (x,y) \to (-x , -y)$; a MIB
is made of $J_1 (x,y) = x^2 $, $J_2 (x,y) = y^2$, and $J_3 (x,y) =
x y$. None of these can be written as a polynomial function of the
others, but $J_1 J_2 = J_3^2$.} and $r$ has the smallest possible value (this value
depends on $G$), we say that they are a {\it minimal integrity
basis (MIB)}. In this case we say that the $\{J_a \}$ are a set of
{\it basic invariants} for $G$. There is obviously some
arbitrarity in the choice of the $J_a$ in a MIB, but the degrees $\{
d_1 , ... , d_k \}$ of $\{J_1 , ... , J_k \}$ are fixed by $G$. (In
mathematical terms, they are determined through the Poincar\'e
series of the graded algebra $P_G$ of $G$-invariant polynomials.)

We will from now on assume we have chosen a MIB, with elements
$\{J_1 , ... , J_k \}$ of degrees $\{d_1 , ... , d_k \}$ in $x$,
say with $d_1 \le d_2 \le ... \le d_k$.

When the elements of a MIB for $G$ are algebraically independent,
we say that the MIB is {\it regular}; if $G$ admits a regular MIB
we say that $G$ is {\it coregular}. 

An algebraic relation between elements $J_\a$ of the MIB is said
to be a relation of the first kind. The algebraic relations among
the $J$ are a set of polynomials in the $\{J_1 , ... , J_r \}$
which are identically zero when seen as polynomials in $x$. If
there are algebraic relations among these, they are called
relations of the second kind, and so on. A theorem by Hilbert
guarantees that the chain of relations has finite maximal length.
(This is the homological dimension of the graded algebra $P_G$
mentioned above.)

In the following we will consider a matrix built with the
gradients of basic invariants, the {\it $\P$-matrix} (Sartori).
This is defined as
$$ \P_{ih} (x) \ := \ \langle \grad J_i (x) , \grad J_h (x) \rangle \eqno(4) $$
with $\langle .,.\rangle$ the scalar product in $\T_* X$.

The gradient of an invariant is necessarily a covariant quantity;
the scalar product of two covariant quantities is an invariant
one, and thus can be expressed again in terms of the basic
invariants. Thus, {\it the $\P$-matrix can always be written in
terms of the basic invariants themselves}.

\section{Geometry of group action}

The use of a MIB allows to introduce a map $\J : x \to \{ J_1 (x) , ... , J_k (x) \}$ from $X$ to a subset $P$ of $\R^k$. If the
MIB is regular, $P = \R^k$, while if the $J_i$ satisfy some
relation then $P \ss \R^k$ is the submanifold satisfying the
corresponding relations. The manifold $P$ is isomorphic to the
orbit space $\Omega = X/G$ (the isomorphism being realized by the
$\J$ map) and provides a more convenient framework to study $\Omega$.

As mentioned above, on physical terms we are mainly interested in
the orbit space up to equivalence of symmetry type. The set of
points in $X$ (of orbits in $\Omega$) with the same symmetry type
will be called a {\it $G$-stratum} in $X$ (a $G$-stratum in
$\Omega$); the $G$-stratum of the point $x$ will be denoted as $\s
(x) \ss X$ (the $G$-stratum of the orbit $\om$ as $\S (\om) \ss
\Om$).\footnote{The notion of stratum was introduced by Whitney in
topology; a {\it stratified manifold} is a set which can be decomposed as the disjoint union of smooth manifolds of different dimensions, the topological (or Whitney) strata: $M = \cup M^k$, with $M^k \ss \pa M^j$ for all $k < j$.}

It results that the $G$-stratification is compatible with the
topological stratification. Indeed $P$ is a semialgebraic (i.e.,
is defined by algebraic equalities and inequalities) stratified
manifold in $\R^k$; the image of any $G$-stratum in $\Om$ belongs
to a single topological stratum in $P$, and topological strata in
$P$ are the union of images of $G$-strata in $\Om$.

Moreover, the subgroup relations correspond to bordering relations
between $G$-strata: if $[G_x] < [G_y]$, then $\s (y) \in \pa \s
(x)$ and (with $\om_x $ the orbit of $x$) $\S (\om_y ) \in \pa \S
(\om_x)$.

There is a stratum, called the {\it principal stratum} $\s_0$,
which corresponds to minimal isotropy, open and dense in $X$;
similarly, the principal stratum $\S_0$ is open and dense in
$\Omega$.

\section{Landau polynomial}
\def\B{\cal B}
\def\L{\cal L}

In the Landau theory of phase transitions (Landau 1937) the state of the system under study is described by a $G$-invariant polynomial $\Phi : X \to \R$ having a critical point in the origin, with at least some of its coefficients -- in particular those controlling stability of the
zero critical point --  depending on external control parameters
(usually, $X = \R^n$ and $G \sse O(n)$; in particular, in solid
state physics $G$ is a crystallographic group). This should be
chosen as the most general $G$-invariant polynomial  of the lowest
degree $\ell$ sufficient to ensure termodynamic stability; in
mathematical terms, this amounts to the requirement that there is
some open set $\B$ containing the origin and such that -- for all
values of the control parameters -- $\grad \Phi$ points inwards at
all points of $\pa \B$ (i.e. $\B$ is invariant under the gradient
flow of $\Phi$). If the polynomials in the MIB are of degree $d_1
\le d_2 ... \le d_r $, then usually $\ell = 2 d_r$.

The $G$-invariance of $\Phi$ and the results recalled above mean
that we can always write it in terms of the polynomials in a MIB
for $G$ as in (3), $\Phi (x) = \Psi [\J (x) ]$.

The discussion of previous sections show that we can study
symmetry breakings for $\Phi : X \to \R$ by studying critical
points of $\Psi : P \to \R$; in other words, {\it Landau
theory can be worked out in the $G$-orbit space $\Om := M / G$}.
The polynomial $\Psi$ -- providing a representation of the Landau
polynomial in the orbit space -- will also be called {\it
Landau-Michel polynomial} \footnote{Louis Michel (1923-1999)
pioneered the use of orbit space techniques in Physics and
Nonlinear Dynamics, originally motivated by the study of hadronic
interactions.}

In this way the evaluation of the map $\Phi : X \to \R$ is in
principles substituted by evaluation of two maps, $\J : X \to P $
and $\Psi : P  \to \R$. However, if -- as in Landau
theory -- we have to consider the most general $G$-invariant
polynomial on $X$, we can just consider the most general
polynomial on $P$.

\section{Critical points of the Landau polynomial and geometry of orbit space}

The $G$-invariance has consequences on the critical points of
$\Phi$. We have already seen one such consequence: critical points
come in $G$-orbits.

This, however, is not all. Indeed, $G$-invariance enforces the
presence of a certain set $\chi (G) \in X$ of critical points, and
conversely if we look for points which are critical under {\it
any} $G$-invariant potential, these are precisely the points in
$\chi (G)$; the critical points on $\chi (G)$ correspond to critical orbits which we call {\it principal critical orbits}.

The set $\chi (G)$ can be determined on the basis of the geometry of the $G$-action.\footnote{A trivial example is provided by $X = \R$ and $G = Z_2$ acting via $g : x \to - x$; any even function has a critical point in zero, and albeit even functions can and in
general will have nonzero critical points, this is the only
critical point common to all the even functions.}  
Indeed (Michel 1971): {\it An orbit $\om$ is a principal critical orbit if and only if it is isolated in its stratum}.

For the linear orthogonal group actions in $\R^n$ often occurring
in Physics, no nonzero point or orbit can be isolated in its
stratum. However, we can quotient out the radial
degeneracy and work on $X = S^{n-1} \ss \R^n$.
In this case, a $G$-orbit $\om_1$ in $S^{n-1}$ which is isolated
in its stratum corresponds to a one-dimensional family $\{\om_r
\}$ of $G$-orbits in $\R^n$, call $X_0$ the corresponding
submanifold in $X$; the gradient of $\Phi$ at $x \in X_0$
points along $\T_x X_0$. We can thus reduce to consider
the restriction $\Phi_0$ of the potential $\Phi$ to $X_0$. (See
also the {\it reduction lemma} of Golubitsky and Stewart in this
context.)

Correspondingly, if $P_0 \ss P$ is the submanifold in $P$
image of $X_0$, i.e. $P_0 = \J (X_0 )$, we can reduce to
consider the restriction $\Psi_0$ of $\Psi$ to $P_0$.

As these become one-dimensional problems, general results are
available. In particular, one can provide general conditions
ensuring the existence of one-dimensional branches of
symmetry-breaking solutions bifurcating from zero along any such
$X_0$ or $P_0$; this is also known as the {\it equivariant
branching lemma} of Cicogna and Vanderbauwhede.

\section{Reduction of the Landau potential}

In realistic problems, $\Phi$ becomes quickly extremely
complicated, i.e. includes a high number of terms and therefore of
coefficients. A thorough study of different symmetry breaking
patterns, i.e. of the symmetry type of minima of $\Phi$ for
different values of these coefficients and of the external control
parameter, is in this case a prohibitive task. It is possible to reduce the generality of the Landau polynomial with no loss of generality for the corresponding physical problem.
Indeed, a change of coordinates in the $X$ space will produce a
formally different -- but obviously equivalent -- Landau
polynomial; it is convenient to use coordinates in which the Landau polynomial is simpler.

A systematic and algorithmic reduction procedure -- based on
perturbative expansion near the origin -- is well known in
dynamical systems theory (Poincar\'e-Birkhoff normal forms), and
can be adapted to the reduction of Landau polynomials.\footnote{An
alternative and more general -- but also much more demanding --
approach is provided by the {\it spectral sequence} approach, also
originating in normal forms theory.}

We work near the origin, so that we can assume $X = \R^n$ (with metric $\eta$), and for simplicity we also take the case where $G$ acts via a linear representation $T_g$.
We consider changes of coordinates of the (Poincar\'e) form
$$ x^i \ = \ y^i \ + \ h^i (y) \ , \eqno(5) $$
generated by a $G$-invariant function $H$: $ h^i (y) = \eta^{ij}
(\pa H (y) / \pa y^j)$; this guarantees that (5) preserves the
$G$-invariance of $\Phi$.
The action of (5) on $\Phi$ can be read from its action on the
basic invariants $J_a$. It results 
$$ \begin{array}{l}
J_a (x) \ = \ J_a (y) \ + \ (\de J_a) (y) \ ; \ \ \ \ \de J_a \ := \
\P_{a b} \, (\pa H / \pa J_b ) \ . \end{array}  \eqno(6) $$

Let us now consider the reduction of an invariant polynomial $\Phi (x) = \Psi (J)$. We
write $D_\a := \pa / \pa J_\a$, and understand summation over
repeated indices is implied. In general, 
$$ \Psi (J) \ \to \ \Psi (J + \de J) \ = \ \Psi (J) + \sum_{\a=1}^r \, {\pa \Psi (J) \over \pa J_\a} \, \de J_\a \ + \ {\rm h.o.t.} \ .  $$

Disregarding higher order terms and using (6) and (4), we get
$$ \de \Psi \ = \ {\pa \Psi \over \pa J_\a} \, \P_{\a \b} \, {\pa H \over \pa J_\b} \ \equiv \ (D_\a \Psi ) \, \P_{\a \b} \, (D_\b H ) \ . \eqno(7) $$

We expand $\Phi$ as a sum of homogeneous polynomials, and write $
\Phi (x) = \sum_{k=0}^\ell \Phi_k (x)$, where $\Phi_k (a x) =
a^{k+1} \Phi_k (x)$. Also, write $\Psi = \sum_k \Psi_k$, where
$\Phi_k (x) := \Psi_k [J(x)]$.

It results that under a change of coordinates (5) generated by $H
= H_m$ homogeneous of degree $m+1$, the terms $\Psi_k$ with $k \le
m$ are not changed, while the terms $\Psi_{m+p}$ change according
to
$$ \Psi_{m+p} \ \to \^\Psi_{m+p} = \Psi_{m+p} + (D_\a \Psi_p ) \P_{\a \b} (D_\b H_m ) \ + \ {\rm h.o.t.} \ . \eqno(8) $$

We can then operate sequentially with $H_m$ of degree
$3,4,...$; at each stage (generator $H_m$) we are not
affecting the terms $\Psi_k$ with $k \le m$. Moreover, we can just
consider (8), as higher order terms are generic and will be taken care of in subsequent steps. (This procedure requires to determine suitable generating functions $H_m$; these are obtained as solutions to {\it homological equations}.)

In the above we disregarded dependence on the control parameters, such as temperature, pression, magnetic field, etc; i.e. we implicitly
considered fixed values for these. However, they have to change
for a phase transition to take place. If we consider a
full range of values -- including in particular critical ones --
for the control parameters, say $\la \in \La$, we should take care
that the concerned quantities and operators are nonsingular
uniformly in $\La$.

This leads to reduction criteria for the Landau and Landau-Michel polynomials (Gufan). Define, for $i=1,...,k$ and with $(.,.)$ the scalar product in $X = \R^n$, the quantities $ U_i (J_1,...,J_k) := ( \pa F / \pa J_s) \, \P_{s i}$. 

\medskip\noindent
{\bf Reduction criterion.} {\it For $\Phi (x) = \Psi(J_1,...,J_k) : \R^n \to \R$ a $G$-invariant potential depending on physical parameters $\la \in \La$, there is a sequence of Poincar\'e changes of coordinates such that $\Phi$ is expressed in the new coordinates $y$ as $\^\Phi (y) = \^\Psi (J)$, where terms which can be written
(up to higher order terms) uniformly in $\La$ as
$ \sum_{\a=1}^k \, Q_\a (J_1 , ... , J_k) \, U_\a (J_1 , ... , J_k )$, 
with $Q_\a$ polynomials in $J_1,...,J_k$ satisfying the
compatibility condition $ (\pa Q_\b / \pa J_\a) = (\pa Q_\a / \pa
J_\b)$, are not present in $\^\Psi$.}

\section{Non-stationary and non-variational problems}

So far we have considered stationary physical states. In some cases, one is not satisfied with such a description, and wants to study time evolution. A model framework for this is provided by the Ginzburg-Landau equation
$$ \xd \ = \ f (x) \eqno(9) $$
where $f = \eta (\grad \Phi) : X \to \T X$ (see above for
notation). In this case, $G$-invariance of $\Phi$ implies
equivariance of (9). More generally, we can consider (9) for an
equivariant smooth $f$ (not necessarily a gradient), i.e. $ f^i (g x) = (Dg)^i_j f^j (x) $.

In this case one shows that
$$ f (x) \ \in \ \T_x \, \s(x) \ , \eqno(10) $$ so that
closure of $G$-strata are dynamically invariant, and the dynamics can
be reduced to them. This is of special interest for the ``most
singular'' strata, i.e. those of lower dimension. The reduction
lemma and the equivariant branching lemma mentioned above also hold (and were originally formulated) in this context.

The relation (10) also implies that one can project the dynamics
(9) in $X$ to a smooth dynamics ${\dot p} = F (p)$ in the orbit
space; this satisfies $F[\J ( x)] = (D \J) [f (x)]$. In the gradient
case this (together with initial conditions) embodies the full
dynamics in $X$, while in the generic case one looses all
information about motions along group orbits (note these correspond
to phonon modes).

An orbit $\om$ isolated in its stratum is still an orbit of fixed
points for any $G$-equivariant dynamics in $X$ in the gradient
case, while in the generic case it corresponds to a fixed point for $F$ and to {\it relative equilibria} (dynamical orbits which belong to a single group orbit) in $X$. In this case, time averages of physical quantities can be $G$-invariant for nontrivial relative equilibria. 

\section{Extensions and physical applications}

We have discussed finite groups symmetry breaking and focused on
polynomial potentials (which can be thought of as Taylor
expansions around critical points). For non finite groups, and in
particular non compact ones, the situation can be considerably
more complicated.

(1) An extension of the theory sketched here is provided by
Palais' theory, and in particular by his {\it symmetric
criticality principle} which applies in Hilbert or Banach spaces
of sections of a fiber bundle satisfying certain conditions. This is specially relevant in connection with field theory and gauge groups.

(2) We focused on the situation discussed in Classical Physics. Finite groups symmetry breaking is of course also relevant in Quantum Mechanics; this is discussed e.g. in the classical books by Weyl and Wigner, and in the review by Michel, Kim, Zak and Zhilinskii.

(3) One speaks of ``explicit symmetry breaking'' when a
non-symmetric perturbation is introduced in a symmetric problem.
In the Hamiltonian case (or in the Lagrangian one for Noether
symmetries), hamiltonian symmetries correspond to conserved
quantities, and non-symmetric perturbations make these become approximate constants of motion.

(4) The symmetry of differential equations -- as well as symmetric
and symmetry-breaking solutions for symmetric equations -- can be
studied in general mathematical terms (see e.g. Olver).

(5) Physical applications of the theory discussed here abound in
the literature, in particular through the Landau theory of phase
transitions. A number of these, together with a deeper discussion
of the underlying theory, is given in the monumental review paper
by Michel, Kim, Zak and Zhilinskii (see ``further reading'').

\section*{Further Reading}

\begin{itemize}

\item Abud M and Sartori G, ``The geometry of spontaneous symmetry
breaking'', {\it Ann. Phys. (N.Y.)} {\bf 150} (1983), 307-372

\item Gaeta G, ``Bifurcation and symmetry breaking'', {\it Phys.
Rep.} {\bf 189} (1990), 1-87

\item Gaeta G, ``Lie-Poincar\'e transformations and a reduction
criterion in Landau theory'', {\it Ann. Phys. (N.Y.)} {\bf 312}
(2004), 511-540

\item Gaeta G and Morando P, ``Michel theory of symmetry breaking and gauge theories'', {\it Ann. Phys. (N.Y.)} {\bf 260} (1997), 149-170

\item Golubitsky M, Stewart I and Schaeffer DG, {\it
Singularities and groups in bifurcation theory}, Springer, New York, 1988

\item Gufan YM, {\it Structural phase transitions} (in russian), Nauka, Moscow 1982

\item Landau LD, ``On the theory of phase transitions I \& II'',
{\it Zh. Eksp. Teor. Fiz.} {\bf 7} (1937), 19 \& 627

\item Landau LD and Lifshitz EM, {\it Statistical Physics},
Pergamon Press, London, 1958

\item Michel L, ``Points critiques des fonctions invariantes sur
une $G$-variet\'e'', {\it Comptes Rendus Acad. Sci. Paris} {\bf
272} (1971), 433-436

\item Michel L, ``Symmetry, defects, and broken symmetry'', {\it
Rev. Mod. Phys.} {\bf 52} (1980), 617-650

\item Michel L, Kim JS, Zak J and Zhilinskii B, ``Symmetry,
invariants, topology''; {\it Physics Reports} {\bf 341} (2004),
1-395

\item Olver PJ, {\it Applications of Lie groups to differential
equations}, Springer, Berlin, 1986

\item Palais RS, ``The principle of symmetric criticality'', {\it Comm. Math. Phys.} {\bf 69} (1979), 19-30

\item Palais RS and  Terng CL, {\it Critical point theory and
submanifold geometry}, Springer-Verlag, Berlin, 1988

\item Sartori G, ``Geometric invariant theory in a
model-independent analysis of spontaneous symmetry and
supersymmetry breaking'', {\it Acta Appl. Math.} {\bf 70} (2002),
183-207

\item Toledano JC and Toledano P, {\it The Landau theory of
phase transitions}, World Scientific, Singapore, 1987

\item Toledano P and Dmitriev V, {\it Reconstructive phase
transitions}, World Scientific, Singapore, 1996

\item Weyl H, {\it The theory of groups and quantum mechanics},
Dover, New York, 1931

\item Wigner EP, {\it Group theory and its application to the
quantum mechanics of atomic spectra}, Academic Press, New
York-London, 1959

\item Winternitz P, ``Lie groups and solutions of nonlinear PDEs'', in {\it Integrable systems, quantum groups, and quantum field theory} (NATO ASI 9009), L.A. Ibort and M.A. Rodriguez eds., Kluwer,
Dordrecht 1993

\end{itemize}

\end{document}